# Emergence and Collapse of Quantum Mechanical Superposition: Orthogonality of Reversible Dynamics and Irreversible Diffusion

Gerhard Grössing, Siegfried Fussy, Johannes Mesa Pascasio, and Herbert Schwabl,

*Austrian Institute for Nonlinear Studies*
Akademiehof
Friedrichstr. 10, A-1010 Vienna, Austria
e-mail: ains@chello.at

**Abstract:** Based on the modelling of quantum systems with the aid of (classical) non-equilibrium thermodynamics, both the emergence and the collapse of the superposition principle are understood within one and the same framework. Both are shown to depend in crucial ways on whether or not an average orthogonality is maintained between reversible Schrödinger dynamics and irreversible processes of diffusion. Moreover, said orthogonality is already in full operation when dealing with a single free Gaussian wave packet. In an application, the quantum mechanical "decay of the wave packet" is shown to simply result from sub-quantum diffusion with a specific diffusivity varying in time due to a particle's changing thermal environment. The exact quantum mechanical trajectory distributions and the velocity field of the Gaussian wave packet, as well as Born's rule, are thus all derived solely from classical physics.

## 1. Introduction

In recent years, a particular perspective has been refined on how to deal with the still unresolved puzzles of quantum physics. This perspective distinguishes between two major classes of approaches to said puzzles [1, 2]. In the first class, the common belief is that quantum theory is exact and only in need of the optimal interpretive framework to cope with apparent paradoxes. This class contains Bohmian mechanics, decoherent histories, many worlds, or quantum theory as information, for example, which are all equivalent with respect to empirical predictions. The second class, however, is characterized by the assumption that quantum theory is not exact, but instead emerges from a still more accurate sub-quantum theory, thereby providing also a clearer picture of how the combination of the deterministic and the probabilistic aspects comes about.

In the present paper, the approach of the second class is followed. It is based on, and further elaborates, the modelling of quantum dynamics via non-equilibrium thermodynamics [3, 4, 5], i.e., a sub-quantum theory out of which ordinary quantum theory is shown to emerge. As the latter is characterized, among others, by the superposition principle, one demands from said sub-quantum theory not only an explanation of how this may come about, but also, under which circumstances it will fail. In other words, the envisaged sub-quantum theory must also provide a framework within which one can understand the "measurement process", or, respectively, what is often called the "collapse of the wave function". In this paper, such a framework shall be provided. It will be based on the demonstration that both the emergence and the collapse of the superposition principle depend in crucial ways on whether or not an (average) orthogonality is maintained between reversible dynamics and irreversible processes of diffusion. Moreover, said orthogonality will be shown here to be already in full operation when dealing with a single Gaussian wave-packet. In this way, the essence of the physics behind quantum mechanical superposition will be shown to be grounded in this average orthogonality condition which is thus the main focus of the present paper.

## 2. Co-existence of reversible Schrödinger dynamics and irreversible diffusion

In references [3] and [4], it is shown how the Schrödinger equation can be derived exactly from an underlying classical level with the aid of non-equilibrium thermodynamics. An earlier derivation of the Schrödinger equation from classical mechanics has been provided by Nelson [6], whose work has found subsequent refinements, e.g., by El Naschie [7], Nottale [8], or Fritsche and Haugk [9]. Apart from open questions w. r. t. nonlocal phenomena in the approaches of these derivations, they are also characterized by perhaps not very "natural" assumptions such as competing diffusion-antidiffusion processes, in order to account for the reversible Schrödinger dynamics. (For alternatives to Nelson-type approaches, which still are comparable in spirit, see the earlier work by Chetaev as recently reviewed in Rusov et al. [10], and a whole series of papers by G. 't Hooft, such as [11], for example.)

However, G. N. Ord has provided a lattice random walk model which in the continuum approximation produces the Schrödinger equation as a *projection* from an ensemble of random walks. (For a review, see [12].) To our knowledge, this is the first application in the literature of the strategy to "leave microscopic irreversibility untouched (keeping the random walk completely intact) and simply look carefully for reversible features which are independent of the intrinsic irreversibility of the full system." [12] In other words, "the fact that the projection is orthogonal to that responsible for diffusion allows the reversible dynamics of Schrödinger's equation to coexist with the irreversible behaviour of particle densities (i.e. diffusion)." [12]

Historically, it had already been Schrödinger himself who pointed out the close resemblance of his time-dependent equation with the classical diffusion equation [13]. This formal analogy, with the equations differing only in that Schrödinger's uses an "imaginary diffusion constant" (i.e., instead of a real-valued one), has been extensively discussed by R. Fürth [14]. In his treatise, much space is devoted to a discussion of the behaviour of Gaussian wave packets, both in classical diffusion and in quantum theory. It is there where one can see very clearly the many similarities, but also the subtle differences between both types of evolutions. Therefore, we shall also in the present paper discuss Gaussian wave packets, to begin with, and we shall see how Ord's strategy will provide a fresh look at the whole topic.

For decades, it had been an unquestioned textbook standard that certain features of quantum theory could not have any equivalent in classical physics, i.e., the Heisenberg uncertainty principle, indeterministic behaviour of a particle despite a deterministic evolution of its statistical ensemble over many runs, nonlocal interaction, tunnelling, or, of course, even a combination of all these. However, this old textbook standard is no longer true, because the whole set of features just mentioned (and more) has recently been proven experimentally to occur in a completely classical system. We are referring to the beautiful series of experiments performed by the group of Yves Couder (see, for example, [15, 16, 17, 18]) using small liquid drops that can be kept bouncing on the surface of a bath of the same fluid for an unlimited time when the substrate oscillates vertically. These "bouncers" can become coupled to the surface waves they generate and thus become "walkers" moving at constant velocity on the liquid surface. A "walker" is defined by a lock-in phenomenon so that the drop falls systematically on the forward front of the wave generated by its previous bouncings. It is thus a "symbiotic" dynamical phenomenon consisting of the moving droplet dressed with the Faraday wave packet it emits. In reference [16], Couder and Fort report on single-particle diffraction and interference of walkers. They show "how this wavelike behaviour of particle trajectories can result from the feedback of a remote sensing of the surrounding world by the waves they emit". The content of this statement is practically identical to a classical model of quantum systems which one of us (G. G.) introduced some time ago, called "Quantum Cybernetics", thereby referring to the mutually causal (i.e., "cybernetic") relationship between wave and particle behaviour. [19]

Of course, the "walkers" of Couder's group, despite showing so many features they have in common with quantum systems, cannot be employed one-to-one as a model for the latter, with the most obvious difference being that quantum systems are not restricted to two-dimensional surfaces. Also, the model of quantum cybernetics lacks specifications necessary for a clear-cut, experimental distinguishability from other approaches. However, along with the understanding of how the Schrödinger equation can be derived via non-equilibrium thermodynamics [3, 4], also the mutual relationship of particle and wave behaviour has become clearer. Just as in the experiments with walkers, there exists an average orthogonality also for particle trajectories and wave fronts in the quantum case. In fact, it lies at the heart of the reasons for the emergence of quantum from sub-quantum behaviour in general, and of the superposition principle in particular, as shall be shown below.

## 3. Dispersion of a free Gaussian wave packet: particle trajectories and velocities from purely classical physics

In the thermodynamic approach to quantum behaviour [3, 4, 5], a particle of energy $E = \hbar\omega$ is characterized by an oscillator of angular frequency $\omega$, which itself is a dissipative system maintained in a non-equilibrium steady-state by a permanent troughput of energy, or heat flow, respectively. The latter is a form of kinetic energy different from the "ordinary" kinetic energy of the particle, as it represents an additional, external contribution to it, like, e.g., from the presence of zero point fluctuations. The total energy of the whole system (i.e., the particle as the "system of interest" in a narrower sense and the heat flow constituting the particle's thermal embedding) is assumed as

$$E_{\text{tot}} = \hbar\omega + \frac{(\delta p)^2}{2m}, \tag{3.1}$$

where $\delta p \coloneqq mu$ is said additional, fluctuating momentum of the particle of mass $m$. The appearance of $\hbar$ may either be taken as the (only) empirical input to our approach [3, 4], or it may be considered as being also due to classical physics. (See [5], and further references therein to a growing literature on the subject.)

For the following, it will be helpful to let ourselves be guided by the picture provided by the "walkers" introduced in the previous Chapter. For, also with a walker one is confronted with a rapidly oscillating object, which itself is guided by an environment that also contributes some fluctuating momentum to the walker's propagation. In fact, the walker is the cause of the waves surrounding the particle, and the detailed structure of the wave configurations influences the walker's path, just as in our thermodynamic approach [3, 4, 5] the particle both absorbs heat from and emits heat into its environment, both cases of which can be described in terms of momentum fluctuations.

Thus, if we imagine the bouncing of a walker in its "fluid" environment, the latter will become "excited" or "heated up" wherever the momentum fluctuations direct the particle to. After some time span (which can be rather short, considering the very rapid oscillations of elementary particles), a whole area of the particle's environment will be coherently heated up in this way. (Considering the electron, for example, the fact that it "bounces" roughly $10^{21}$ times per second, with each bounce eventually providing a slight displacement from the original path's momentum, one can thus understand the "area filling" capacity of any quantum path whose fractal dimension was shown to be equal to $2$. [20])

Now, let us assume we have a source of identical particles, which are prepared in such a way that each one ideally has an initial (classical) velocity $\mathbf{v}$. Even if we let them emerge one at a time only, say, from an aperture with unsharp edges (thus avoiding diffraction effects to good approximation), the probability density $P$ will be a Gaussian one. This comes along with a heat distribution generated by the oscillating ("bouncing") particle(s), with a maximum at the center of the aperture $\mathbf{x}_0 = \mathbf{v}t$. So, we have, in one dimension for simplicity, the corresponding solution of the heat equation,

$$P(x,t) = \frac{1}{\sqrt{2\pi}\sigma} e^{-\frac{(x-x_0)^2}{2\sigma^2}}, \qquad (3.2)$$

with the usual variance $\sigma^2 = \overline{(\Delta x)^2} = \overline{(x-x_0)^2}$, where we shall choose $x_0(t=0) = 0$. Note that from Eq. (3.1) one has for the *averages* over particle positions and fluctuations (as represented via the probability density $P$)

$$\overline{E_{\text{tot}}} = \overline{\hbar\omega} + \frac{\overline{(\delta p)^2}}{2m} = \text{ const.,} \qquad (3.3)$$

with the mean values (generally defined in $n$–dimensional configuration space)

$$\overline{(\delta p)^2} := \int P(\delta p)^2 \, d^n x. \qquad (3.4)$$

As opposed to Eq. (3.1), where $\delta p$ can take on an arbitrary value such that $E_{\text{tot}}$ is generally variable, equation (3.3) is a statement of total average energy conservation, i.e., holding for all times $t$. This means that in Eq. (3.3), a variation in $\delta p$ implies a varying "particle energy" $\hbar\omega$, and vice versa, such that each of the summands on the right hand side for itself is not conserved. In fact, as shall be detailed below, there will generally be an exchange of momentum between the two terms providing a net balance

$$m\delta\text{v} - m\delta u = 0, \qquad (3.5)$$

where $\delta\text{v}$ describes a change in the "convective" velocity $\text{v}$ paralleled by the "diffusive" momentum fluctuation $\delta(\delta p) := m\delta u$ in the thermal environment.

As elaborated in references [3] and [4], once Eq. (3.1) is assumed, considerations based on Boltzmann's relation between action and angular frequency of an oscillator provide, without any further reference to quantum theory, that

$$\delta p = mu := \hbar k_u = -\frac{\hbar}{2}\nabla \ln P. \qquad (3.6)$$

Now we make use of one out of a whole series of practical identities, which Garbaczewski has collected in [21]. (These identities hold true on general information theoretic grounds and are thus not bound to quantum mechanical issues.) Said identity, which can easily be checked by integration, is given by

$$\overline{(\nabla \ln P)^2} = -\overline{\nabla^2 \ln P}. \qquad (3.7)$$

In a further step, we now introduce a way to prepare for an Ord-type of projection as mentioned in the previous Chapter, i.e., to cut out a "slice of time" from an otherwise irreversible evolution as given by the assumed diffusion process. To do this, we shall first combine Equations (3.6) and (3.7), and shall then insert (3.2) for the initial time, $t=0$. As from (3.3) one has that $\frac{\partial}{\partial t}\overline{E_{tot}}=0$, and thus also $\delta\overline{E_{tot}}(t)-\delta\overline{E_{tot}}(0)=0$, and as only the kinetic energy varies, one obtains $\delta\overline{E_{kin}}(t)=\delta\overline{E_{kin}}(0)=$ const.. Then, with the Gaussian (3.2), this provides an expression for the averaged fluctuating kinetic energy, or heat, of a particle and its surroundings,

$$\delta\overline{E_{kin}}(t) = \frac{m}{2}\overline{(\delta v)^2} + \frac{m}{2}\overline{u^2} = \frac{m}{2}\overline{(\delta v)^2} + \frac{\hbar^2}{8m\sigma^2} = $$
$$= \delta\overline{E_{kin}}(0) = 0 + \frac{m}{2}\overline{u^2}\big|_{t=0} = \frac{\hbar^2}{8m\sigma_0^2} =: \frac{m}{2}u_0^2. \qquad(3.8)$$

Eq. (3.8) is an expression of the fact that at the time $t=0$ the system is known to be in the prepared state whose fluctuating kinetic energy term is solely determined by the initial value $\sigma_0$, whereas for later times $t$ it decomposes into the term representing the particle's changed kinetic energy and the term including $\sigma(t)$. As the kinetic energy term of the particle increases, the convective velocity becomes $v(t) \to v(t) + \delta v$ for $t>0$, and, correspondingly, $u(t) \to u(t) - \delta u$ for $t>0$. In other words, one can decompose said term into its initial $(t=0)$ value and a subtracted fluctuating kinetic energy term, respectively, i.e.,

$$\frac{\hbar^2}{8m\sigma^2} = \frac{m}{2}\overline{u^2} = \frac{m}{2}u_0^2 - \frac{m}{2}\overline{(\delta u)^2}, \qquad(3.9)$$

where the last term on the right hand side is identical to $\frac{m}{2}\overline{(\delta v)^2}$ in order to fulfil Eq. (3.8), and also in agreement with Eq. (3.5).

From Eqs. (3.8) and (3.9) one derives *minimal uncertainty relations* for all $t$, i.e.,

$$\Delta p \cdot \sqrt{\sigma^2} = \Delta p \cdot \Delta x = \frac{\hbar}{2}, \text{ where } \Delta p := \sqrt{\overline{(\delta p)^2}} = m\sqrt{\overline{u^2}},$$

and, particularly, (3.10)

$$\Delta p_0 \cdot \sqrt{\sigma_0^2} = (\Delta p \cdot \Delta x)_{t=0} = \frac{\hbar}{2}, \text{ where } \Delta p_0 := \sqrt{\overline{(\delta p)^2}}|_{t=0} = m u_0.$$

Moreover, with the "diffusion constant"

$$D := \hbar/2m \tag{3.11}$$

Eq. (3.8) provides an expression for the initial velocity fluctuation,

$$u_0 = \frac{D}{\sigma_0}. \tag{3.12}$$

Let us now consider the emergence of "well ordered" diffusion waves out of the "erratic", Brownian-type diffusions of myriads of single sub-quantum particles through their thermal environments. Being swept along with a diffusion wave, with initial $(t=0)$ location $x(0)$ and diffusion velocity $u$, a quantum particle's distance to the heat accumulation's center $x_0$ at time $t$ will be

$$x(t) = x(0) + ut, \tag{3.13}$$

such that one obtains the r.m.s. of (3.13) as

$$\int x^2 P(x,t) dx = \int (x-x_0)^2 P(x,0) dx +$$
$$2\int (x-x_0) u(x,t) \; t \; P(x,t) dx + \int u^2(x,t) \; t^2 P(x,t) dx,$$
or briefly,
$$\overline{x^2}|_t = \overline{(\Delta x)^2}|_{t=0} + 2 \; \overline{\Delta x \cdot u(x,t)} \; t + \overline{u^2(x,t)} \; t^2. \tag{3.14}$$

Now we introduce the *central argument of the present paper*: we assume, as an emerging result out of the statistics of a vast number of diffusion processes, the complete statistical independence of the velocities $u$ and $v$, and thus also of $u$ and the positions $\Delta x$ $(= x$ for $x_0 := 0)$:

$$\overline{xu} = \overline{vu}\, t = 0. \tag{3.15}$$

This is justified considering the statistics of huge numbers, millions of millions of diffusive sub-quantum Brownian motions, which are supposed to bring forth the emergence of said larger-scale collective phenomenon, i.e., the diffusion wave fields as solutions to the heat equation [4]. (In our associative picture, these are nothing but the analogy to the walkers' Faraday waves emitted with some fixed frequency.) In other words, Eq. (3.13) represents the effect of collectively "smoothing out" the "erratic" processes of individual Brownian motions. Thereby, the mean convective and diffusion velocities must be unbiased (lest one introduces new physics), and thus linearly uncorrelated. (Note that it was exactly the corresponding average orthogonality of momentum and momentum changes which has led to a first new derivation of the Schrödinger equation [22], as well as the subsequent one based on non-equilibrium thermodynamics [3, 4].)

Therefore, with the thus introduced Ord-type projection, i.e., the orthogonality of classical (convective) momentum on one hand, and its associated diffusive momentum on the other, one gets rid of the term linear in $t$ in Eq.(3.14), and thus of irreversibility, and one obtains

$$\overline{x^2} = \overline{x^2}\big|_{t=0} + \overline{u^2} t^2. \tag{3.16}$$

Eq. (3.16) is the result obtained for the "pure" emergent diffusive motion as given by (3.13). However, in a further step we now take into account the small momentum fluctuations $m\delta u$ (which we have discussed above w.r.t. (3.9)), providing an altered convective velocity $v \to v + \delta v(t)$, and thus an additional displacement $\delta x = |\delta u| t = |\delta v| t$, i.e., as soon as $t > 0$. Therefore, in Eq. (3.13) one now must decompose $u(t)$ into its initial value $u_0$ and a fluctuating contribution $\delta u(t)$, respectively. Unless some thermal equilibrium were reached, the latter is typically given off from the "heated" thermal bath to the particle of velocity v,

$$u(t) = u_0 - \delta u(t), \quad (3.17)$$

which is in accordance with (3.9)

As opposed to (3.13), Eq. (3.17) now provides the particle's total displacement

$$x(t) + \delta x(t) = x(0) + ut = x(0) + (u_0 - \delta u)t. \quad (3.18)$$

Squaring (3.18) provides

$$x^2(t) + 2x(t)\delta x(t) + (\delta x(t))^2 = x^2(0) + u_0^2 t^2 - 2u_0 \delta u t + (\delta u)^2 t^2. \quad (3.19)$$

Since $(\delta x)^2 = (\delta v)^2 t^2 = (\delta u)^2 t^2$, one obtains in accordance with Eq. (3.5) that the last terms on the l.h.s. and on the r.h.s. of (3.19), respectively, cancel each other out. Moreover, as the product terms in (3.19) are subject to the average orthogonality condition, one obtains through averaging over positions and fluctuations that

$$\overline{x^2} = \overline{x^2}\big|_{t=0} + u_0^2 t^2. \quad (3.20)$$

Inserting (3.12) into (3.20) for the particular case that $\overline{x^2} = \overline{v^2} t^2 \equiv \sigma^2$ (i.e., $\overline{x^2}\big|_{t=0} \equiv \sigma_0^2$), provides for the time evolution of the wave packet's variance

$$\boxed{\sigma^2 = \sigma_0^2 \left(1 + \frac{D^2 t^2}{\sigma_0^4}\right).} \quad (3.21)$$

The quadratic time-dependence of the variance $\sigma^2$ is remarkable insofar as in ordinary diffusion processes the scenario is different. There, with the Gaussian distribution being a solution of the heat equation, for purely Brownian motion the variance grows only linearly with time, i.e., as described by the familiar relation

$$\overline{x^2} = \overline{x^2}\big|_{t=0} + 2Dt. \tag{3.22}$$

However, as we have seen, the momentum exchange between the particle and its environment is characterized by both a changing velocity and by a changing thermal environment of the particle, i.e., also by a changing diffusivity. Therefore, Eq. (3.22) must be modified to allow for a time-dependent diffusivity. An ansatz, which can easily be shown to comply with the heat equation and its solutions (3.2), is given by the linear $t$- dependence of the form

$$D(t) = u_0^2 t = \frac{D^2}{\sigma_0^2} t = \frac{\hbar^2}{4m^2 \sigma_0^2} t, \tag{3.23}$$

(As will be rigorously shown in a forthcoming paper, (3.23) is the *only possible* time-dependent diffusivity in our scenario.) Moreover, in order to find agreement with an overall Brownian-type of motion, one has to be careful not to deal in (3.23) with too short time intervals of the order $t \approx 1/\omega$. In fact, if one introduces a time-averaged diffusivity

$$\langle D(t) \rangle := \frac{1}{t} \int_0^t D(t') dt' = \frac{u_0^2}{2} t = \frac{D(t)}{2}, \tag{3.24}$$

one immediately obtains the linear-in-time Brownian relation

$$\boxed{\overline{x^2} = \overline{x^2}\big|_{t=0} + 2\langle D(t) \rangle t \quad \text{and} \quad \sigma^2 = \sigma_0^2 + 2\langle D(t) \rangle t,} \tag{3.25}$$

which is, however, also in accordance with the $t^2$ – dependence of Eq. (3.21). Note that the diffusivity's rate of change is a constant,

$$\frac{dD(t)}{dt} = \frac{D^2}{\sigma_0^2} = u_0^2 = \text{const.}, \tag{3.26}$$

such that it is determined only by the initial r.m.s. distribution $\sigma_0$. In other words, the smaller the initial $\sigma_0$, the faster $D(t)$ will change.

With the square root of (3.21),

$$\sigma = \sigma_0 \sqrt{1 + \frac{D^2 t^2}{\sigma_0^4}} \qquad (3.27)$$

we note that $\sigma/\sigma_0$ is a spreading ratio for the wave packet independent of $x$. This functional relationship is thus not only valid for the particular point $x(t) = \sigma(t)$, but for all $x$ of the Gaussian. Therefore, one can generalize (3.27) for all $x$, i.e.,

$$x(t) = x(0)\frac{\sigma}{\sigma_0}, \text{ where } \frac{\sigma}{\sigma_0} = \sqrt{1 + \frac{D^2 t^2}{\sigma_0^4}} \;. \qquad (3.28)$$

In other words, one derives also the time-invariant ratio

$$\frac{x(t)}{\sigma} = \frac{x(0)}{\sigma_0} = \text{ const.} \qquad (3.29)$$

Now we remind ourselves that we deal with a particle of velocity $v = p/m$ immersed in a wave-like thermal bath that permanently provides some momentum fluctuations $\delta p$. The latter are reflected in Eq. (3.27) via the r.m.s. deviation $\sigma(t)$ from the usual classical path. In other words, one has to do with a wave packet with an overall uniform motion given by $v$, where the position $x_0 = vt$ moves like a free classical particle. As the packet spreads according to Eq.(3.27), $x(t) = \sigma(t)$ describes the motion of a point of this packet that was initially at $x(0) = \sigma_0$. Depending on whether initially $x(0) > \sigma_0$ or $x(0) < \sigma_0$, then, respectively, said spreading happens faster or slower than that for $x(0) = \sigma_0$. In our picture, this is easy to understand. For a particle exactly at the center of the packet $(x_0)$, the momentum contributions from the "heated up" environment on average cancel each other for symmetry reasons. However, the further off a particle is from that center, the stronger this symmetry will be broken, i.e., leading to a position-dependent net acceleration or deceleration, respectively, or, in effect, to the "decay of the wave packet". Moreover, also the appearance of the time-dependent diffusivity $D(t)$ is straightforward in our model. Essentially, the "decay of the wave packet" simply results from sub-quantum diffusion with a diffusivity varying in time due to the particle's changing thermal environment: as the heat initially concentrated in a narrow spatial domain gets gradually dispersed, so must the diffusivity of the medium change accordingly.

In conclusion, then, one obtains with Eqs. (3.28) and (3.11) for the "smoothed out" *trajectories* (i.e., those averaged over a very large number of Brownian motions)

$$\boxed{x_{tot}(t) = vt + x(t) = vt + x(0)\frac{\sigma}{\sigma_0} = vt + x(0)\sqrt{1 + \frac{\hbar^2 t^2}{4m^2 \sigma_0^4}}} \quad . \tag{3.30}$$

Fig. 1 provides a graphic representation of Eq. (3.30) for an exemplary set of trajectories. Moreover, one can now also calculate the average total velocity,

$$v_{tot}(t) = \frac{dx_{tot}(t)}{dt} = v(t) + \frac{dx(t)}{dt}. \tag{3.31}$$

Thus, with (3.28), one obtains the *average total velocity field of a Gaussian wave packet* as

$$\boxed{v_{tot}(t) = v(t) + \left[x_{tot}(t) - vt\right]\frac{\hbar^2}{4m^2}\frac{t}{\sigma^2 \sigma_0^2}.} \tag{3.32}$$

Next to the fundamental relations (3.25), Equations (3.30) and (3.32) are the main results of this part of the paper. They provide the trajectory distributions and the velocity field of a Gaussian wave packet as derived solely from classical physics. Note that the trajectories are not the "real" ones, but only represent the averaged

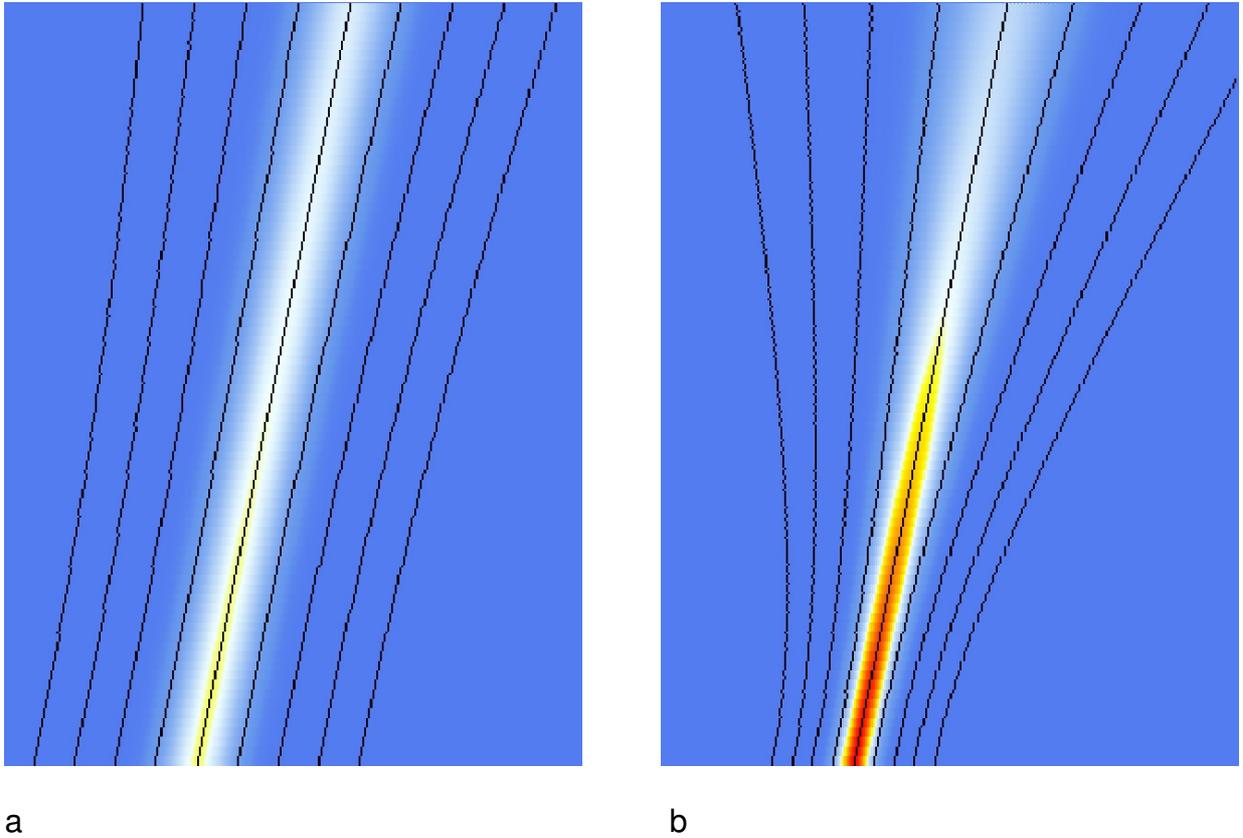

**Fig. 1 a** and **b : Dispersion of a free Gaussian wave packet.** Considering the particles of a source as oscillating "bouncers", they can be shown to "heat up" their (generally non-local) environment in such a way that the particles leaving the source (and thus becoming "walkers") are guided through the thus created thermal "landscape". In the Figures, the classically simulated evolution of exemplary *averaged* trajectories is shown (i.e., averaged over many single trajectories of Brownian-type motions). These trajectories are thus no "real" trajectories, but they only represent the averaged behaviour of a statistical ensemble. The results are in full agreement with quantum theory, and in particular with Bohmian trajectories. This is so despite the fact that no quantum mechanics is used in the calculations (i.e., neither a quantum mechanical wave function, nor a guiding wave equation, nor a quantum potential), but purely classical physics.

The Figures display a simulation with coupled map lattices of classical diffusion and a time-dependent diffusivity as given by Eq. (3.23). Two examples are shown, with different half-widths of the initial Gaussian distribution, respectively: (1+1)-dimensional space-time diagrams (time axis from bottom to top), with the intensity field and nine exemplary averaged trajectories, in agreement with Eq. (3.30). In Fig 1a, the initial half width $\sigma_0$ is twice as large as in Fig. 1b. Note that the narrower the Gaussian distribution is concentrated initially around the central position, the more the thus "stored" heat energy tends to push trajectories apart.

behaviour of a statistical ensemble. The results are in full concordance with quantum theory, and in particular with Bohmian trajectories. (For a comparison with the latter, see, for example, [23], or the Figures for the Gaussian wave packet example in [24], which are in excellent agreement with our Fig. 1.) This is so despite the fact that no quantum mechanics has been used yet, i.e., neither a quantum mechanical wave function, or the Schrödinger equation, respectively, nor a guiding wave equation, nor a quantum potential.

Implicitly, of course, one can easily find the connections to the rhetoric of (Bohmian or other) quantum mechanics. As for the Bohmian case, one just needs to consider the expression for the quantum potential,

$$U(x,t) = -\frac{\hbar^2}{2m}\frac{\nabla^2 \sqrt{P}}{\sqrt{P}}. \tag{3.33}$$

Then one has, again with the help of the general relation (3.7),

$$-\overline{\frac{\nabla^2 \sqrt{P}}{\sqrt{P}}} = \frac{1}{4}\overline{(\nabla \ln P)^2} = -\frac{1}{4}\overline{\nabla^2 \ln P}, \tag{3.34}$$

and thus obtains from Eq. (3.2) for $t=0$ the time-independent expression for the *average quantum potential* as

$$\overline{U(x,0)} = -\frac{\hbar^2}{8m}\overline{\nabla^2 \ln P}\Big|_{t=0} = \frac{\hbar^2}{8m\sigma_0^2}. \tag{3.35}$$

The expression (3.35) is identical to the one we obtained on the r.h.s. of (3.8), such that we find that the energy conservation law (3.3) can be rewritten as

$$\overline{E_{\text{tot}}} = \hbar\omega + \overline{U} = \text{const.}, \tag{3.36}$$

i.e., where

$$\overline{U}(x,t) = \frac{\overline{(\delta p)^2}}{2m}. \tag{3.37}$$

Still, it is rather remarkable that the results presented above can be arrived at without even referring to (Bohmian or other) quantum mechanics. However, let us now see how a "translation" of the present formalism into that of ordinary quantum mechanics can be accomplished.

## 4. The superposition principle and Born's rule from classical physics

### 4.1. The "translation scheme"

The "translation" between the language of classical physics employed so far in this paper on one hand, and traditional quantum theory on the other, can easily be established. As emphasized in Chapter 3 as the central argument of the present paper, our main focus is on the *average orthogonality* between reversible physics (i.e., as represented by the classical velocity $\mathbf{v}$ of the center of a Gaussian wave packet, for example) and irreversible diffusion due to a "heated" environment (i.e., as represented by the additional velocity $u$). This average orthogonality is explicitly described by Eq. (3.15) in classical terms for our example of the Gaussian, but it can be re-written (returning now to a description with three spatial dimensions) and generalized for all $\mathbf{v}$ and $u$ in the following way:

$$\overline{\mathbf{vu}} = 0, \text{ i.e., } \overline{\mathbf{v}_{\text{tot}}^2} = \frac{\overline{\mathbf{x}_{\text{tot}}^2}}{t^2} = \overline{\mathbf{v}^2} + \frac{\overline{\mathbf{x}^2}}{t^2} =: \overline{\mathbf{v}^2} + \overline{\mathbf{u}^2} = \overline{|\mathbf{v} + i\mathbf{u}|^2}. \tag{4.1.1}$$

Now, we already know that **u** is given via Eq. (3.6) by

$$\mathbf{u} = -\frac{\hbar}{2m}\left(\frac{\nabla P}{P}\right), \quad (4.1.2)$$

and the classical momentum is usually given by the gradient of the action $S$, such that

$$\mathbf{v} = \frac{\nabla S}{m}. \quad (4.1.3)$$

Thus, we have that the average total momentum squared can be written as

$$\overline{\mathbf{p}_{tot}^2} = m^2 \overline{\mathbf{v}_{tot}^2} = \overline{\left|\nabla S - i\frac{\hbar}{2}\left(\frac{\nabla P}{P}\right)\right|^2} = \hbar^2\left[\overline{\left(\frac{\nabla S}{\hbar}\right)^2} + \frac{1}{4}\overline{\left(\frac{\nabla P}{P}\right)^2}\right]. \quad (4.1.4)$$

Considering now that the intensity of a wave (packet) is generally represented via the amplitude $R(\mathbf{x},t)$ as

$$P(\mathbf{x},t) = R^2(\mathbf{x},t), \quad (4.1.5)$$

one obtains

$$\overline{p_{tot}^2} = \hbar^2 \overline{k_{tot}^2} = \hbar^2\left[\overline{\left(\frac{\nabla S}{\hbar}\right)^2} + \overline{\left(\frac{\nabla R}{R}\right)^2}\right] =: \hbar^2 \overline{\left|\frac{\nabla \psi}{\psi}\right|^2}, \quad (4.1.6)$$

where a "compactification" is achieved by the introduction of a "wave function" $\psi$, defined as

$$\psi(x,t) = R(x,t) e^{i\frac{S(x,t)}{\hbar}}. \quad (4.1.7)$$

In other words, a re-formulation of the classical total momentum $\mathbf{p}_{tot}$ as a complex-valued one,

$$\mathbf{p}_{tot} = m(\mathbf{v} + i\mathbf{u}) =: \hbar(\mathbf{k} + i\mathbf{k_u}) = \hbar \mathbf{k}_{tot}, \quad (4.1.8)$$

reads in terms of the quantum mechanical wave function as

$$\hbar \mathbf{k}_{tot} = -i\hbar\left(\frac{\nabla \psi}{\psi}\right) = \nabla S - i\hbar\left(\frac{\nabla R}{R}\right), \quad (4.1.9)$$

from which the average length (squared) of the vector, Eq. (4.1.6), can be obtained. As is well known, the introduction of $\psi$ thus provides a linearization of an otherwise more complicated set of coupled differential equations [22]. Let us now see how the quantum mechanical superposition principle and Born's rule can be formulated with the aid of our classical physics approach. To do so, we firstly consider the description of the physics if two alternative paths are present (like, e.g., in interferometry), and then, secondly, what happens when two consecutive paths for one particle are given, or (as it will turn out) equivalently, an anti-correlated two-particle system.

### 4.2. Two alternative paths, A or B

To guide our imagination, let us again refer to the walkers discussed in the previous Chapter. As was shown in experiment [16], the bouncing "particle" can be sent through a two-slit system such that the "particle" itself just passes one slit, whereas its accompanying Faraday waves pass through both slits, interfere behind the slits, and guide the walker to a screen where, eventually, an interference pattern is registered. In analogy to this scenario, we now discuss our system of "particle plus wave-like thermal bath" along two possible paths, $A$ and $B$. Firstly, we note that conservation of the total momentum demands that

$$|\mathbf{p_{tot}}| = \hbar |\mathbf{k_{tot}}| = \hbar |\mathbf{k}_A| = \hbar |\mathbf{k}_B|. \tag{4.2.1}$$

Secondly, with this momentum conservation, $k_{tot} = k_A = k_B$, in classical physics two overlapping waves with amplitudes $R_A$ and $R_B$, respectively, provide

$$R(x,t)\mathbf{k}_{tot} = R_A(x,t)\mathbf{k}_A + R_B(x,t)\mathbf{k}_B. \tag{4.2.2}$$

Now consider the average squared momentum as given in (4.1.6). With the aid of Eqns. (4.1.5) and (4.2.2), one can write for the average explicitly

$$\overline{\hbar^2 k_{tot}^2} = \int d^n x \, P \, \hbar^2 k_{tot}^2$$
$$= \hbar^2 \int d^n x \, \left(\sqrt{P}\mathbf{k}_{tot}\right)^2 = \hbar^2 \int d^n x \, \left(\sqrt{P_A}\mathbf{k}_A + \sqrt{P_B}\mathbf{k}_B\right)^2 \qquad (4.2.3)$$
$$= \hbar^2 \int d^n x \, \left(R\mathbf{k}_{tot}\right)^2 = \hbar^2 \int d^n x \, \left(R_A\mathbf{k}_A + R_B\mathbf{k}_B\right)^2.$$

Division by $\hbar^2 k_{tot}^2$ then provides, with $\hat{k}_i$ ($i = A$ or $B$) denoting the unit vectors, the normalized integral of the intensity, or probability density $R^2(x,t)$,

$$P_{tot} = \int d^n x P = \int d^n x R^2 = \int d^n x \left(R_A \hat{\mathbf{k}}_A + R_B \hat{\mathbf{k}}_B\right)^2. \qquad (4.2.4)$$

Actually, when taking the square roots in (4.2.4), one has to note that generally $P$ can be either $(+R)^2$ or $(-R)^2$, indicating that for waves which are anti-symmetric around their origin (e.g., a particle source), the amplitude summation will turn into an amplitude subtraction. Empirically, of course, this applies exactly to fermions, but apart from this, no additional assumption is necessary that would be of a purely quantum mechanical (i.e., as opposed to classical) nature. So, with this possibility as a caveat, we shall continue our discussion, thereby restricting ourselves to the option of equation (4.2.2).

Now, according to (4.1.9),

$$P_{tot} =: \overline{|\psi_{tot}|^2} = \frac{\overline{|\nabla \psi_{tot}|^2}}{k_{tot}^2}. \qquad (4.2.5)$$

There thus remains to be shown the following: In order to agree with our classical equation (4.2.4), it must hold in quantum mechanical terms that

$$\psi_{tot} = \psi_A + \psi_B. \qquad (4.2.6)$$

So, we combine Eqns. (4.1.7), (4.2.6), and (4.2.5), and substitute the division by $k_{tot}^2$ by a general normalization factor $N$. Then, one can rewrite (4.2.5) as

$$P_{\text{tot}} = \overline{\left|\sum_a \frac{1}{\sqrt{N}}\left[\frac{\nabla S_a}{\hbar} - i\left(\frac{\nabla R_a}{R_a}\right)\right]\psi_a\right|^2} =: \overline{\left|\sum_a c_a \psi_a\right|^2}. \tag{4.2.7}$$

Note that we have thus introduced the complex-valued coefficients

$$c_a = \frac{1}{\sqrt{N}}\left[\frac{\nabla S_a}{\hbar} - i\left(\frac{\nabla R_a}{R_a}\right)\right] =: \frac{1}{\sqrt{N}}\widetilde{c}_a, \tag{4.2.8}$$

with the normalization

$$\sum_a |c_a|^2 = 1 \iff \sum_a |\widetilde{c}_a|^2 = N = k_{\text{tot}}^2. \tag{4.2.9}$$

For $a = 1$, one can easily confirm Eq. (4.2.9):

$$|\widetilde{c}_a| = \left|\frac{\nabla S}{\hbar} - i\left(\frac{\nabla R}{R}\right)\right| = \left|\frac{\nabla \psi}{\psi}\right| = k_{\text{tot}}. \tag{4.2.10}$$

In order to check for $a = 2$, a more lengthy calculation is required. One obtains after some steps that

$$\left|\frac{\nabla \psi_{\text{tot}}}{\psi_{\text{tot}}}\right|^2 = \left|\frac{\nabla(\psi_A + \psi_B)}{\psi_A + \psi_B}\right|^2 = \frac{1}{R_A^2 + R_B^2 + 2|R_A||R_B|\cos(\delta S/\hbar)} \cdot$$

$$\left\{\begin{array}{l} (\nabla R_A)^2 + (\nabla R_B)^2 + \left(\frac{\nabla S_A}{\hbar}\right)^2 R_A^2 + \left(\frac{\nabla S_B}{\hbar}\right)^2 R_B^2 + 2\nabla R_A \cdot \nabla R_B \cos(\delta S/\hbar) + \\ 2R_A \nabla R_A \cdot \nabla S_A \sin(S_A/\hbar) + 2R_B \nabla R_A \cdot \nabla S_B \sin(S_A/\hbar) + \\ 2R_A \nabla R_B \cdot \nabla S_A \sin(S_B/\hbar) + 2R_B \nabla R_B \cdot \nabla S_B \sin(S_B/\hbar) + \frac{2}{\hbar} R_A R_B \nabla S_A \cdot \nabla S_B \end{array}\right\}. \tag{4.2.11}$$

Now follows the important step in our calculation, as we remind ourselves of the central importance of the average orthogonality of $\nabla R_i$ and $\nabla S_i$, respectively, with $i = A$ or $B$. It holds, as can easily be seen, even for the cases where the indices $i$ are not identical: because of the spherically symmetrical distributions of the wave vectors, and thus of the average $\overline{\nabla R_i}$, *any average product* $\overline{\nabla R_{i_1} \cdot \nabla S_{i_2}}$, i.e., with $i_1 = i_2$ or $i_1 \neq i_2$, vanishes due to orthogonality. Thus, the corresponding terms in (4.2.11) are to be deleted when calculating the averages. In fact, one finally obtains from Eq.

(4.2.11), even when introducing different weights $\alpha$ on different paths (such as $R_B = \alpha R_A$) that *generally*

$$\overline{\left|\frac{\nabla \psi_{tot}}{\psi_{tot}}\right|^2} = \overline{\left(\frac{\nabla R}{R}\right)^2} + \overline{\left(\frac{\nabla S}{\hbar}\right)^2} = k_{tot}^{\ 2}, \qquad (4.2.12)$$

with $\overline{\left|\frac{\nabla R_i}{R_i}\right|^2} =: \overline{\left|\frac{\nabla R}{R}\right|^2}$, $\overline{|\nabla S_i|^2} =: \overline{|\nabla S|^2}$. This confirms Eq. (4.2.9) for $a = 2$.

We thus see that the conservation of the (squared) momentum, $k_{tot}^{\ 2}$, is only guaranteed when the averaging procedures necessary to simplify Eq. (4.2.11) are fully in operation. In other words, then, we have shown that there exists an *equivalence* between two "finely tuned" calculatory schemes, i.e., the implementation of the *average orthogonality of classical particle momenta and their wave-related fluctuations on one hand*, and *the superposition principle on the other*. Both can unambiguously be "translated" into each other, as has just been shown.

Moreover, Eqns. (4.2.7) through (4.2.9) imply that we have derived from classical physics the following statement: when, in quantum mechanical terms, a system is described by the total wave function,

$$\psi_{tot}(x,t) = \sum_a c_a \psi_a(x,t), \qquad (4.2.13)$$

the probability of finding the result $a$ is given by

$$\int P_a(x,t) d^n x = |c_a|^2. \qquad (4.2.14)$$

This is *Born's rule*, which has been proven here for $a = 1$ or $2$, but can by induction be extended to $a = n$ alternative possibilities.

Let us now turn to applications. In fact, considering an example from interferometry with two possible alternatives, our classical formula (4.2.4) is immediately applied. Using (4.1.7) and the orthogonality of $\mathbf{k}$ and $\mathbf{k}_u := \delta\mathbf{k}$, one obtains with $k_A = k_B = k$:

$$P = R_A^2 \hat{\mathbf{k}}_A^2 + R_B^2 \hat{\mathbf{k}}_B^2 + 2R_A R_B \cos(\hat{\mathbf{k}}_A \cdot \hat{\mathbf{k}}_B)$$
$$= R_A^2 + R_B^2 + 2R_A R_B \cos\Delta\phi, \quad (4.2.15)$$
$$\text{where } \Delta\phi = \hat{\mathbf{k}}_A \cdot \hat{\mathbf{k}}_B = (S_A - S_B)/\hbar.$$

This agrees exactly with the quantum mechanical result, providing now also an example of a calculation without wave functions for the *double slit*: Normalization provides with (4.2.2) that

$$1 \cdot \mathbf{k}_{tot} = \frac{1}{\sqrt{2}} 2\mathbf{k}, \text{ and thus } k_{tot}^2 \equiv 2k^2,$$

and therefore $N = \frac{1}{2}$. Thus, $P = \frac{1}{2}(R_A \hat{\mathbf{k}}_A + R_B \hat{\mathbf{k}}_B)^2$, and with $R_A = R_B = \frac{1}{\sqrt{2}}$ one obtains, solely on our "classical" basis, the correct result for the intensity distribution on a screen registering an interference pattern:

$$P = \frac{1}{4}(\hat{\mathbf{k}}_A + \hat{\mathbf{k}}_B)^2 = \frac{1}{2}(1 + \cos\Delta\phi). \quad (4.2.16)$$

### 4.3. Two consecutive paths for one particle and the anti-correlated two-particle system

As a further application of our classical approach, we now consider two consecutive paths for one particle and the anti-correlated two-particle system.

In both cases, it holds that $\mathbf{k}_{(u)} = \mathbf{k}_{(u)1} + \mathbf{k}_{(u)2}$, and also that $\overline{k_{tot}^2} = \overline{k^2} + \overline{k_u^2}$. Thus,

$$P(x_1, x_2)\overline{k_{tot}^2} = N\left[\overline{(R_1\mathbf{k}_1 + R_2\mathbf{k}_2)^2} + \overline{(R_1\mathbf{k}_{u1} + R_2\mathbf{k}_{u2})^2}\right], \tag{4.3.1}$$

where the indices 1 and 2 can either denote two consecutive paths for one particle, or two particles in an anti-correlated system (i.e., with opposite momenta). Choosing for simplicity $R_1 = R_2 = R$, we obtain that

$$\begin{aligned} P(x_1, x_2) &= \\ &= N\frac{R^2}{k_{tot}^2}\left[\overline{(\mathbf{k}_1 + \mathbf{k}_{u1})^2} + \overline{(\mathbf{k}_2 + \mathbf{k}_{u2})^2} + 2\overline{(\mathbf{k}_1 \cdot \mathbf{k}_2 + \mathbf{k}_{u1} \cdot \mathbf{k}_{u2})}\right] \\ &= N\frac{R^2}{k_{tot}^2}\left[k_{tot,1}^2 + k_{tot,2}^2 + 2\mathbf{k}_{tot,1} \cdot \mathbf{k}_{tot,2}\right] = NR^2\left[\hat{k}_{tot,1} + \hat{k}_{tot,2}\right]^2. \end{aligned} \tag{4.3.2}$$

With $R^2 = \frac{1}{2}$, $k_{tot,1} = k_{tot,2} = k$ and $k_{tot}^2 = \left(\frac{1}{\sqrt{2}} 2\mathbf{k}\right)^2 = 2k^2$ and thus $N = \frac{1}{2}$, one obtains for the example of the anti-correlated two-particle system

$$P(x_1, x_2) = \frac{1}{4}[1 + 1 + 2\cos\delta\phi], \tag{4.3.3}$$

$$\begin{aligned} \delta\phi &= \frac{\Delta S_{tot}}{\hbar} = \Delta\mathbf{x} \cdot \Delta\mathbf{k}_{tot} = (\mathbf{x}_1 - \mathbf{x}_2) \cdot (\mathbf{k}_1 - \mathbf{k}_2) \\ &= [(\mathbf{x}_1 - \mathbf{x}_0) - (\mathbf{x}_2 - \mathbf{x}_0)] \cdot (\mathbf{k}_1 - \mathbf{k}_2) =: 2\mathbf{r} \cdot (\mathbf{k}_1 - \mathbf{k}_2). \end{aligned}$$

Thus,

$$P(x_1, x_2) = \frac{1}{2}[1 + \cos(2\Delta\phi)] = \cos^2\Delta\phi = \cos^2[\mathbf{r} \cdot (\mathbf{k}_1 - \mathbf{k}_2)]. \tag{4.3.4}$$

Note that this result again agrees exactly with the corresponding calculation in orthodox quantum theory, like, e.g., for the calculation of the intensity distributions in [25]. One particular feature of (4.3.4) is given by the decidedly nonlocal correlation for said distributions. We can now also note that our "classical" expression

$$\overline{k_{tot}^2} = \overline{\left(\frac{\nabla S_1}{\hbar} + \frac{\nabla S_2}{\hbar}\right)^2} + \overline{\left(\frac{\nabla R_1}{R_1} + \frac{\nabla R_2}{R_2}\right)^2}$$

$$= \overline{\left[\nabla\left(\ln R_1 + \ln R_2\right)\right]^2} + \overline{\left[\frac{1}{\hbar}\nabla\left(S_1 + S_2\right)\right]^2} \quad (4.3.5)$$

$$= \overline{\left(\frac{\nabla R_1 R_2}{R_1 R_2}\right)^2} + \overline{\left(\frac{\nabla\left(S_1 + S_2\right)}{\hbar}\right)^2}.$$

implies that quantum mechanically

$$\psi_{tot} = \psi_1 \psi_2. \quad (4.3.6)$$

For the proof, note that $\psi_1\psi_2 = R_1 R_2 e^{\frac{i}{\hbar}(S_1+S_2)}$ and thus:

$$\overline{\left|\frac{\nabla(\psi_1\psi_2)}{\psi_1\psi_2}\right|^2} = \overline{\left(\frac{\nabla(R_1 R_2)}{R_1 R_2}\right)^2} + \overline{\left(\frac{\nabla(S_1+S_2)}{\hbar}\right)^2} = \overline{k_{tot}^2}.$$

As, according to our transformation law (4.2.5), $P_{tot} = \dfrac{\overline{|\nabla\psi_{tot}|^2}}{\overline{k_{tot}^2}}$, we obtain

$$P_{tot}(\mathbf{x}_1, \mathbf{x}_2) = \overline{|\psi_1\psi_2|^2}. \quad (4.3.7)$$

Finally, we note that any path can in principle be considered to be decomposable into two sub-paths, such that the same procedure applies as shown here. This means that the induction from two to $n$ consecutive steps is straightforward. As a similar argument also holds for the addition of $n$ alternative paths, we have shown that Born's rule can be understood completely on the basis of our "classical" approach. In particular, the linearity of the quantum mechanical superposition principle is explained by classical relations of the type (4.2.2). One reason, therefore, why so far no nonlinear modification to the Schrödinger equation could be observed experimentally is given by the circumstance that some fluctuations, often considered as the sources of the hypothesized nonlinearities, are already *constitutive* for the validity of the linear laws like (4.2.2): the average wave vectors $\mathbf{k}_{tot}$ already contain momentum fluctuation components, *and* are also subject to the average orthogonality condition.

## 5. Conclusions and Perspectives: Towards a classical theory of the collapse of quantum mechanical superposition

In this paper we have investigated some consequences of modelling quantum systems with "walker"-type oscillations in the thermal bath of a vacuum structured by zero point fluctuations. We have restricted ourselves to the non-relativistic case, although a generalization to the relativistic one should be feasible. In fact, there exists a very interesting relativistic description of quantum systems by Baker-Jarvis and Kabos [26], in which they clearly distinguish between "particle" and wave contributions to generally complex-valued momenta, similarly as discussed in this paper. That is, one can work out a quantum dynamics distinguishing particle momenta $\hbar k$ and their accompanying waves' contributions $\hbar \delta k$ such that the relativistic energy-momentum law reads

$$\hbar^2 \omega^2 + \hbar^2 c^2 \overline{(\delta k)^2} = \hbar^2 \left( \omega_0^2 + c^2 \overline{k_{tot}^2} \right), \text{ with } \overline{k_{tot}^2} = \overline{(k + \delta k)^2}, \qquad (5.1)$$

where we have used the average orthogonality condition $\overline{k \cdot \delta k} = 0$. Having worked here with the non-relativistic variant of the total energy $E_{tot}$, i.e.,

$$E_{tot} = \hbar \omega + \frac{\hbar^2 \overline{(\delta k)^2}}{2m} = \hbar \omega_0 + \frac{\hbar^2 \overline{k^2}}{2m} + \frac{\hbar^2 \overline{(\delta k)^2}}{2m}, \qquad (5.2)$$

one can make the following observation.

Relating the momentum fluctuation $\delta p = mu = \hbar \delta k$ solely to the emergent wave behaviour, or, the oscillator's basic dynamics, respectively, such that

$$u = \omega r, \tag{5.3}$$

where

$$r = 2\sigma_0 = \sqrt{\frac{2D}{\omega}} \tag{5.4}$$

is the usual diffusion length, one obtains with Eq. (3.10) that

$$m\omega \frac{r^2}{2} = \frac{\hbar}{2}, \tag{5.5}$$

and thus

$$\frac{mu^2}{2} = \frac{\hbar^2 (\delta k)^2}{2m} = m\frac{\omega^2 r^2}{2} = \frac{\hbar \omega}{2}. \tag{5.6}$$

So, we see that without the momentum fluctuations, a quantum system's "total energy" $\hbar \omega$ is given by only the first two terms on the right hand side of Eq. (5.2). However, inclusion of the thermal environment provides the full quantum version of the total energy with the zero-point fluctuations. The corresponding additional term is identical with the average quantum potential (3.37),

$$\overline{U} = \frac{\hbar \omega}{2}. \tag{5.7}$$

Moreover, note that with (5.3) one can express the frequency $\omega$ as

$$\omega = \frac{u}{r} = \frac{D}{2\sigma_0^2} = \frac{\hbar}{4m\sigma_0^2}, \tag{5.8}$$

such that one can also rewrite Eq. (3.30) as

$$x_{\text{tot}}(t) = vt + x(0)\sqrt{1 + 4\omega^2 t^2}. \tag{5.9}$$

Here we just observe that Eq. (5.8) gives us a clear statement about a "quantum walker's" frequency. Considering, on the one hand, the smallest possible diffusion length as

$$r = \frac{1}{\delta k} = \frac{\lambda_C}{2\pi} = \frac{\hbar}{m_0 c}, \qquad (5.10)$$

with $\lambda_C$ being the Compton wavelength, one obtains that $u = \frac{2D}{r} = c$, and

$$\hbar\omega = m_0 c^2, \qquad (5.11)$$

thus providing the familiar zitterbewegung frequency, e.g., for the electron, $\omega = \omega_{ZB} \approx 10^{21} Hz$. On the other hand, if one prepares a system with broader distributions such as Gaussians, Eq. (5.8) shows that the larger one chooses $\sigma_0$ to be, the smaller the frequency $\omega$ becomes. This is exactly what one would expect from a "walker", i.e., a maximal hitting (or bouncing) frequency "on the spot" (of size $\sim \lambda_C$), and an ever decreasing hitting frequency $\omega$ for ever larger $\sigma_0$.

We thus arrive at a clear picture also of the ontological status of the various entities in our sub-quantum model. In contrast, on the one hand, to hydrodynamical models of the sub-quantum regime, which provide no clear statement of how an individual particle is to be distinguished from the "rest" of the "flow" of probability distributions, we have in our model the definite movement of a localized entity, i.e., a "particle" (which may well be the nonlinear part of a wave), surrounded by the "flow" of its wave-like environment. The latter is described via non-equilibrium thermodynamics, which is considered in purely classical terms. This, on the other hand, is in stark contrast to Bohmian mechanics, where it is the quantum mechanical wave function $\psi$ that is supposed to be "real" and thus to "influence" the motion of actual particle configurations. We thus claim that our model has a much more clear-cut position to offer with respect to the "reality" of quantum systems, in that it can be completely described in terms of (modern) classical physics, i.e., without a "$\psi$ that falls from the sky".

Finally, then, one can also derive from our model some consequences for the understanding of the "measurement problem". The origin of the latter is given by the unpredictability of individual measurement results despite a deterministic law of (unitary) evolution, once a well-prepared state is known at some initial time $t_0$. Most approaches to the problem maintain that quantum theory should apply to both the particle passing through an experimental setup and to the measuring device. That is, the final state at time $t$ is then given by a unitary evolution $U = \exp(-iHt)$ applied to the state at $t_0$, thus describing a superposition and not the stochastic patterns of mutually exclusive measurement results. However, in Chapter 4 we have found an *equivalence between the superposition principle* on the one hand *and the average orthogonality of particle momenta and wave-related fluctuating momenta, respectively,* on the other. A violation of said orthogonality has been shown to result in a violation of the *conservation of the average momentum* $\overline{p}$. For example, average orthogonality violation in Eq. (4.2.11) would in general result in a violation of the momentum conservation (4.2.12). In other words, then, said violation of average momentum conservation would immediately provide also a violation of the superposition principle, i.e., a deviation from unitary evolution.

The actual measurement process can thus be understood as a process of symmetry breaking, ultimately resulting in energy/momentum transfer from the particle to the detecting apparatus. Said momentum transfer breaks the symmetry of the superposition principle (i.e., where *all* possible measurement outcomes are represented via coherent addition of corresponding probability amplitudes) and transforms unitary evolution into a non-unitary one. It is thus beyond the domain of application of the Schrödinger equation and must generally be looked-for in the context of a non-equilibrium thermodynamics as, for example, discussed in [3, 4]. In reference [3], a "vacuum fluctuation theorem" has been presented as an extension of the model discussed here, which applies to integrable non-conservative systems and is of interest for our present purposes. Considering that to some quantum system a non-vanishing average work $\overline{\Delta W}$ is applied, or, contrariwise, the quantum system provides some work $\overline{\Delta W}$ to its environment, one has for the corresponding vacuum thermodynamics that probabilities $\mathrm{p}$ for heat dissipation $(A)$ or absorption $(-A)$ are related by

$$\frac{\mathrm{p}\left(\frac{1}{t}\frac{\overline{\Delta W}}{kT} = \frac{1}{t}\frac{\overline{\delta U}}{\hbar\omega} = A\right)}{\mathrm{p}\left(\frac{1}{t}\frac{\overline{\Delta W}}{kT} = \frac{1}{t}\frac{\overline{\delta U}}{\hbar\omega} = -A\right)} = e^{At} = e^{\frac{\overline{\delta U}}{\hbar\omega}}, \quad (5.12)$$

with $\overline{\delta U}$ being a difference in the average quantum potential $\overline{U}$. [3] This provides an "external", and possibly non-local, momentum fluctuation $\delta \mathbf{p}_{\text{ext}}$, i.e., in addition to the usual momentum fluctuations $\delta \mathbf{p}$ discussed in this paper so far,

$$\delta \mathbf{p}_{\text{ext}} = \frac{1}{2}\nabla\left(\frac{\overline{\delta U}}{\omega}\right). \quad (5.13)$$

Comparing with Eq. (3.6), this provides the total momentum fluctuation as [3]

$$\delta \mathbf{p}_{\text{tot}} = \delta \mathbf{p} + \delta \mathbf{p}_{\text{ext}} = -\frac{\hbar}{2}\nabla \ln\{P + \mathrm{p}(-A) - \mathrm{p}(A)\}. \quad (5.14)$$

So, one understands how a non-vanishing gradient of fluctuations in the average quantum potential, (5.13), can account for the symmetry breaking which violates time-reversible, unitary evolution and the superposition principle. Dissipation of kinetic energy with probability $\mathrm{p}(A)$ thus provides the increase in $\delta\mathbf{p}_{\mathrm{tot}}$ that potentially completes the "measurement process". In this way, one sees how irreversibility comes back into the game on the observational level, i.e., as soon as the average orthogonality between unitary Schrödinger dynamics and irreversible diffusion processes is discarded.